\documentstyle[12pt,epsf,aaspp4]{article}

\def\kms{{\rm\,km/s}}
\def\msun{{\rm\,M_\odot}}

\def\vol#1  {{{#1}{\rm,}\ }}

\def\etal{et al.\ }

\newcount\refno
\newcount\rfno
\def\eq{$^{\the\refno\ }$\advance\refno by 1}
\def\ad{\advance\rfno by 1}

\def\clock{\count0=\time \divide\count0 by 60
        \count1=\count0 \multiply\count1 by -60
\advance\count1 by
\time

\number\count0:\ifnum\count1<10{0\number\count1}
\else\number\count1\fi}

\def\kms{\rm km/s}

\def\Gcm2{\rm G~cm^2}

\def\beq{\begin{equation}}
\def\eeq{\end{equation}}

\def \date         {\ifcase\month \message{zero}
\or
                       January \or February \or
March \or April \or May \or June
                       \or July \or
                       August \or September \or
October \or November
\or
                       December \fi
                       \space\number\day,
\number\year}

\begin{document}

\title{Cosmological Mestel Disks and The Rossby Vortex
Instability: The Origin of Supermassive Black Holes }
\author{Stirling A. Colgate\altaffilmark{1},  Renyue
Cen\altaffilmark{2}, Hui
Li\altaffilmark{1}.  Nathan Currier\altaffilmark{1},
and Michael S. Warren\altaffilmark{1}}
\altaffiltext{1} {MS 227, Los Alamos National Laboratory,
P.O. Box 1663, MS B227, Los Alamos, NM 87545; colgate@lanl.gov.}

\altaffiltext{2} {Princeton University
Observatory,  Princeton University, Peyton Hall, Ivy Lane, Princeton, NJ
08544; cen@astro.princeton.edu.}
\received{\date}
\accepted{ }

\begin{abstract} A scenario is put forth for the
formation of supermassive black holes at the centers of
galaxies. It depends upon the formation of a
Mestel disk with a flat rotation curve,  $M_{<r}
\propto r$  and $ \Sigma \propto 1/r $.
Such disks could form from the collapse of   uniformly  rotating,
isolated,  gaseous clouds, either proto-galactic, galaxy-mass
damped  Ly$\alpha$ clouds or the gas that survives galaxy
mergers.    We propose that in either case the  disk will be
unstable to the Rossby  vortex instability (RVI).   This instability
grows from any large, steep pressure gradient in an optically thick
disk.  Such pressure gradients either occur adjacent to
compact objects or could be triggered by  heating from
individual supernovae in  and around the disk.  Upon excitation, the
RVI  transports angular momentum outward, accreting nearly all mass
within the initiation radius. We have calculated that in very thin
disks,  the non-linear vortices initiated by the RVI can
transport angular momentum  far more efficiently than turbulence.
Compared to a viscosity-based Shakura-Sunyaev disk, the RVI
transports angular momentum out to a much larger radius, so more
mass is accreted  into the central black hole.  A typical
galaxy rotational velocity is
$v_{rot} = 200$ km s$^{-1}$, and the critical column density,
necessary to initiate the RVI is
$\Sigma_{CCD} \simeq 100$ g cm$^{-2}$.  For
$M_{<r} = 2 \pi r^2
\Sigma$, we have
$r_{CCD} = v_{rot}^2/ ( 2 \pi \Sigma_{CCD} G) $,   and the   mass
accreted becomes
$M_{BH} = v_{rot}^4/(2 \pi \Sigma_{CCD} G^2)   =
3 \times 10^7 M_{\odot}$. Both the black hole mass
$M_{BH}$ and its
   $v_{rot}^4$ dependence are in good agreement
with  recent observations,  because  $v_{rot}  =  \sqrt 3 \;
\sigma_c$, where $\sigma_{c}$ is the velocity dispersion of the
bulge at the radius of mutual  contact.

\end{abstract}

\keywords{accretion disks --- black hole physics --- galaxies:formation
galaxies:kinematics and dynamics --- hydrodynamics --- instabilities }

\section{INTRODUCTION}

There is mounting evidence that non active
galaxies, including our own, harbor central supermassive black holes (SMBHs)
with masses from  $10^6$ to $10^9\msun$ (Richstone \etal 1998;
Tremaine \etal 2002).
Furthermore, observations indicate a close correlation
between the SMBH mass and the host galaxy bulge mass,
and an even tighter correlation between the SMBH mass and
the bulge velocity dispersion, $\sigma_c$.
How and when the SMBHs form has remained unknown, although there
are many possibilities (Rees 1984; Ostriker 2000;
Adams, Graff, \& Richstone 2001; Loeb 1993; Eisenstein \& Loeb
1995; Umemura, Loeb, \& Turner 1993).

Mestel (1963) developed a model of cloud collapse which preserves
specific angular momentum.  His spherical cloud collapsed to a disk
with a nearly flat rotation curve, but a uniformly rotating inviscid
cylindrical cloud should collapse to a disk with a perfect flat 
rotation curve.  This  mass distribution,
$M_{<r} \propto r$, should extend inward  to  where pressure forces 
become important,
e.g.,   the  mass enclosed within a solar radius $r_{\odot}$
becomes $M = 0.2 M_{\odot}$.  The Helmholtz cooling time at every
radius is $\sim 10^4$ yr, which is short compared to the BH 
formation time of $10^8$ yr, so the disk is thin. 
Turbulence has been  invoked to
form a Shakura-Sunyaev (SS) disk at larger radii. However, for a
$10^8$ $M_{\odot}$ BH, an SS  accretion disk would become
subject to the self-gravitating instability (SGI)   at
$\sim<0.01 \; \mbox{pc}$ (Begelman, Frank, \& Shlosman
1989;  Shlosman et al. 1990; Goodman 2003; Shakura \&
Sunyaev 1973, Gammie 2001).  At 0.08 pc, the disk  would
have more mass than the BH because of
the excessive SS column density required to
transport  angular momentum at a rate corresponding to nearly the
Eddington luminosity. Hence, even  if the  SGI turbulence is
excited and  maintained, the disk mass is still  greatly
excessive.  The large-scale vortices of the Rossby vortex instability 
(RVI) disk are one solution to this problem.

  An Eddington limit (1 $M_{\odot}$ yr$^{-1}$), Thompson opacity
SS disk surrounding a central mass of  $M_{BH} \simeq  10^8
M_{\odot}$ results in $H_{SS}/r \simeq 10^{-4}$ (Shakura \&
Sunyaev 1973; Frank, King, \& Raine 1985; Goodman 2003; Papaloizou
\& Pringle 1984). Therefore, with  the critical column density (CCD)
necessary to initiate the RVI at
$r_{CCD}  \simeq 10^3 r_{SS} \simeq 10 $ pc,  the mass enclosed  
becomes $\pi  r^2 \Sigma_{CCD} \simeq 10^8 M_{\odot}$.    All
turbulent viscosity mechanisms, no matter how robust,  will lead to
a disk mass too large for the  disk to be consistent with the
observed mass, growth time, and luminosity of active galactic nuclei.
The RVI disk does  not rely on turbulence but on vortices  with
transport lengths of $\it{l} \sim r/10$, which is
much greater than   $\it{l} \sim H \sim r / 10^4$ for the SS disk.

\section{THE COLLAPSE OF A GASEOUS CLOUD AND FORMATION OF A MESTEL DISK}
\label{sec:collapse}

A self-gravitating cloud  is parameterized by its mass,
$M_{\rm v} = (4 \pi/3) \rho {r_{\rm v}}^3$, initial (virial)
radius, $r_{\rm v}$, initial  (virial) velocity
dispersion, $v_{\rm v} = (M_{\rm v} G /3 r_{\rm v})^{1/2}$,
and the dimensionless spin parameter,
\begin{equation}
\lambda \equiv J|E|^{1/2}G^{-1}M_{\rm v}^{-5/2}.
\end{equation}
\noindent Here J is  the total angular
momentum, and $|E| = \int  M_{<r} (G / r) dM$ describes the
potential energy  (Peebles 1969).   For a lone test particle  with
tangential velocity
$v_{\rm T}$ as it orbits a fixed mass, equation (1) gives
$\lambda = v_{\rm T}/ v_{\rm K} = 1$, where $v_{\rm K}$ is
the Keplerian velocity.  We are interested in  the
final  collapsed  state of a
rotating   cloud.  The mass at its outer
boundary will  form the outer boundary of the
equilibrium disk.  However, simulations
calculate the mass average of  $\lambda$,  $\lambda_{av}$, not
its value at the outer surface.  For our purposes we renormalize
$\lambda$ to $\lambda_{eq}$ so that  a particle in a Keplerian
orbit tracing an object's equator has  $\lambda_{eq} =  1$.

We also greatly simplify Mestel's analysis by  treating a sphere
as a right circular cylinder with  the sphere's mass,
uniform density, and equal total angular momentum. The
cylindrical approximation  introduces  only a 3\% change in radius,
yet the expected variation in precollapse cloud shapes is very
much greater. The advantage of the cylindrical approximation is
that in Mestel's model it forms a flat rotation disk, because  each
cylindrical shell at
$r_i$  can be treated as a separate test particle that
collapses  to its own Keplerian radius,
$r_{K}$.   For a cylinder rotating as a solid body, the
specific  angular momentum starts as
$j(r_{i})\propto r_{i}^2$ and
$M(<r_{i}) \propto r_{i}^2$,  so $j(r_{i})\propto
M(<r_{i})$. After collapse,
$j(r_{K})\propto (M r_{K})^{1/2}$, so $M \propto (M
r_{K})^{1/2}$,   and one obtains $M \propto r_K$: i.e.,  a flat
rotation curve as seen in the calculations of Bullock
\etal (2001) and Cen \etal (in preparation).

Using our equatorial  definition of  $\lambda_{eq}$,  a typical
gaseous cloud, collapsing  with no transport of angular momentum,
becomes rotationally supported at
$r_0 \sim \lambda_{eq}^2 r_{cloud}  \simeq 1 $ kpc
starting  from a typical cloud with $r_{cloud}= 300$ pc,
$v_{\rm v}\sim 100\kms$, and $M_v \simeq 10^{12} M_{\odot}$
of dark and ($\sim 15\%$) baryonic matter.  A typical value of
$\lambda_{av}$ for a cloud formed  in cosmological structure
formation  is
$0.05$ (Warren \etal 1992; Steinmetz
\& Bartelmann 1995; Cen \etal (in preparation)). The radius of support is about
$1/10$ of a typical L* galaxy  radius of 10 kpc, a well-recognized
dilema, leaving in doubt
this explanation of BH formation based on the transition from
inviscid to vortex flow at the critical column density.  
However, we see two factors
that should reduce this large discrepancy: (1) the conversion  from
$\lambda_{eq}$ to
$\lambda_{av}$  and (2) the reduction of interior
mass as the dark matter and baryonic matter separate during
the  collapse.

   For a sphere, the mass averaged
$\lambda_{av} = 0.31$ if the equator rotates at the Keplerian
velocity.  This is because an integration of  $\lambda$
combines  the moment of inertia of a sphere,
$I = (2/5 M R^2 \omega)$, the self-energy $|E| = (3/5) M^2
G/R$, $\omega = |E|^{1/2} /R$, and
$v_{sph} = \omega R$. Thus the disk radius  predicted  by
$r_0$ above should be
larger by  $(1/0.31)^2 = 10.4$, giving agreement with observed disk
size because of this correction alone.  However,  a further
correction may be made for the decrease of total interior mass as
the baryonic matter falls deep  inside  the dark matter
Navarro-Frenk-White distribution, $\rho(r) / \rho_{crit} =
\delta_c / ((r/r_s) \cdot (1+r/r_s)^2)$.  For  $r << r_s$,   the
interior mass of dark matter decreases as $M_{dk} \propto r^2$,
but the baryonic Mestel disk follows
$M_{bary} \propto r$, giving $\rho_{dark} << \rho_{CCD}$.
We  note that the collapsed radius depends sensitively on the
initial spin parameter
$\lambda_{eq}$, which implies that the total amount of gas and
hence the SMBH masses could have a large dispersion,
corresponding to the dispersion in  $\lambda_{eq}$.

\subsection{Making a Gaseous Disk by Galaxy Mergers}

In addition to the collapse of a single gas cloud as described
above,  in  galaxy-galaxy mergers the gas fraction  of the
progenitors should   collide and make  a uniformly
rotating gas cloud. When the two clouds collide, they produce a gas
pressure corresponding to the kinetic energy of collision.  This
occurs  before  their subsequent collapse by cooling.  The
$\lambda_{eq}$  of the now combined cloud  will depend
sensitively upon whether the collision was prograde or retrograde
(Van den Bosch
\etal 2002), but in either case a merged cloud will rotate as
uniform body, collapse, and form a disk with nearly flat rotation.

A typical bulge might have $10^{10}$ $M_{\odot}$ or $10\%$ of the
mass of a massive galaxy.   However,
because we expect mergers to take place after initial 
galaxy formation, the gas fraction should be $10^{9}$ $M_{\odot}$ 
or $\sim 10\%$  of the bulge stars.   The outer
boundary of a typical bulge merges with the flat disk at $\sim
500 $ pc  where
the velocity is Keplerian.   The relaxation of a uniformly
rotating  bulge of stars and   gas should leave the merged
stars  in inclined  orbits,  thus forming the visible  bulge.
The gas will collapse,  shock,  and
cool just as in the collapse of the original   cloud,
forming a sub-disk of  $10^9$ $M_{\odot}$ with a flat rotation
curve  within $\sim 500$ pc. Barnes \&
Hernquist (1991) and Mihos \& Hernquist (1996) have
simulated such a merger  producing a  more compact
mass  inside
$\sim 200$ pc.

\section{THE RVI IN A MESTEL DISK}
\label{sec:rvi}

The RVI is a global hydrodynamic
instability in thin disks, excited by a radial extremum in an
entropy-modified version of potential vorticity (Lovelace
\etal 1999; Li \etal 2000). Its nonlinear evolution has been
studied using global two-dimensional 
hydrodynamic  simulations in which the flow
of matter through the vortices is  well illustrated in Fig. 6 of Li
\etal (2001). One may rightfully ask why the SGI  is not
sufficient to transport the angular momentum  as Gammie (2002) has
discussed, but  if the RVI works, then the strong shocks  may
disperse slowly growing self-gravity perturbations.  In fact, the
SGI might even be useful, because it could help initiate the RVI.
The RVI, once initiated, has been shown to efficiently
transport angular momentum outward, especially by  the large-scale
vortices that it produces.  In order to initially excite the 
instability,
there must be a  large enough local pressure gradient associated
with the potential vorticity extremum.
The RVI can be initiated  provided the pressure gradient condition
is met.   Also, the column density of the disk
has to be high enough so that locally generated heat can be
trapped for at least several orbits before radiative cooling
damps the fluid velocity stresses. This thickness condition is
actually quite  generic for many hydrodynamic
instabilities in disks since they rely on  pressure forces
to drive fluid motions.  The only hurdle is the need for
positively correlated Reynolds stresses for outward  transport of
angular momentum.

The growth rate of the RVI depends on the sound
speed,  but if the collapse were to reach equilibrium at a
temperature of
$\sim 10$ K, the sound speed would be
$c_s \simeq 10^{-3} v_{\rm v}$.  This low sound speed would
combine with an orbit time of
$t_{orbit} \simeq 3 \times 10^5$ yr at 10 pc to yield an RVI
growth time of $\sim 10^9$ yr, which is unrealistically
long.  However, during the initial collapse there is  both a
time   and  a radius at which the
column density is great  enough to trap a major fraction of
the heat of collapse,  allowing for rapid growth of the RVI.
This  would ideally
lead  to a sound speed
$c_{s} = (\gamma_{adiab} - 1)^{1/2} \simeq v_{\rm v}/3 $, where
$\gamma_{adiab} =  4/3$.  A disk with $\Sigma = 100$  g cm$^{-2}$ 
collapsing at 1 pc radius   is likely  to be initially
supported by radiation pressure when collapsing through a  disk
height $H \approx 0.1 r$, with a corresponding   $T \simeq (\rho
c_s^2 /a)^{1/4}  = ((v_{\rm v}/10)^2 \Sigma /aH )^{1/4} = 740$
K.  At this temperature the opacity is  of the order of
$\kappa \simeq 1$ cm$^2$ g$^{-1}$,  and the radiation pressure
is  $\simeq 10 \times nkT $.  Consequently, the cooling time
becomes $\tau_{cool} \simeq H (\Sigma / \kappa)/ (c/3)
\simeq 10^{-2} \Omega ^{-1}$, far too fast for the RVI to be
initiated. However, as the disk cools to $100$ K,  the height
shrinks to $H \approx 0.01 r$, the disk  becomes supported by
matter pressure below $T = 340$ K, and with an opacity of
$\kappa \simeq 1$ cm$^2$ g$^{-1}$,  the cooling time increases
to $\tau_{cool} \simeq H (\Sigma / \kappa)/ (c/3)) (T C_{p}
\rho/ aT^4)   \simeq  0.4 \Omega ^{-1}$.  If we consider
somewhat greater thickness at correspondingly smaller $r$,
then the cooling time scales as
$\tau_{cool} \simeq H  (\Sigma / \kappa)/ (c/3)) (T C_{p} \rho/
aT^4) \simeq (r / r_{100})^{3}
\Omega ^{-1} = (\Sigma / \Sigma_{100})^3   \Omega ^{-1}$.  Thus
there is a thickness  $100 \kappa < \Sigma_{RVI} < 1000 \kappa$
\, g cm$^{-2}$ where the RVI should maintain itself with an
instability growth time $\propto (r / r_{100})^{3}$, which is a
fraction of the   SMBH formation time.  Three-dimensional
simulations are planned, a laboratory experiment has been
proposed (Colgate \& Buchler 2000), based on evidence that  Rossby
vortices dominate the transport of angular momentum in Earth's
atmosphere.

\subsection{Angular momentum transport and triggering of
the RVI}

   Li \etal (2001) have studied various
initial disk pressure profiles that are unstable to the RVI in a
disk where $\Sigma$ is uniform. (Since the initiating vortices
have radii $r_v  \ll r$, we expect the Mestel disk, with $\Sigma
\propto  r$, to behave similarly.) The RVI can have
a growth rate of
$\sim 0.1-0.3
\Omega$ when the initiating scale length associated with the
local pressure gradient $P_s/(dP_s/dr)$ is approximately
$3H$  (Fig. 6, Li \etal 2001). Large pressure gradients produce
strong nonlinear vortices with strong shocks. In these calculations
the shocks reach pressure ratios of $P_s > 3 \rho c_s^2$   with
$c_s = 0.1 v_{Kep}$.  The
pressure contrast of such shocks is large compared to that
expected in strong turbulence where
$c_{turb} \leq 0.3 c_s$,  so
$P_{shock}/P_{turb} \simeq 3 c_s^2/c_{turb}^2 \sim 30$.  Thus
we postulate  that the RVI  generates the necessary pressure
contrast in  shocks  such that the large-scale
$l \simeq r/10$ transport  of angular momentum  persists
regardless of how thin  the disk  is. The  torque  produced by
the RVI  becomes $H_{RVI} \simeq (r^2\cdot H)
(P_s/3)$. The ratio
of torque produced by the RVI  to that produced by turbulent
viscosity  is
$H_{RVI}/H_{turb} \simeq 0.03 P_s/(\rho
c_s^2)(H/r) $, where  $H_{turb}  \simeq (r^2\cdot H) (\rho
c_s^2/9)(H/r)$.   Both azimuthal pressure gradients
$P_s/3$  and $\rho (c_s/3)^2 \cdot (H/r)$ act
with a torque arm r over an area $Hr$,  but the turbulent stress is
smaller because the turbulent eddy scale is limited to H (Pringle
1981).  Thus in the special case of a massive BH's
accretion disk where  $H/r \simeq 10^{-4}$, the
RVI torque may be greater than the SS turbulent torque by a
factor of $\sim 10^5$.

During the cloud collapse, we expect the SMBH to be seeded by a
BH from a giant  central star.  Each new stage of the star's
evolution  causes a steep pressure gradient at the inner boundary
of the disk. If the
RVI starts on the inside, it must progress to the outside
where all the mass resides.   There should also be a significant
supernova rate, perhaps one per 100 yr per $10^8 M_{\odot}$ or
roughly  $10^6$ supernovae in the formation time of the SMBH.   A
single supernova  of
$10^{51}$ ergs near the disk is sufficient to heat  a  local area
of the disk ($\sim  0.01 \pi r^2 \Sigma$)  and raise the sound
speed to $c_{s,SN} \simeq 0.1 v_{\phi}$,  high enough to initiate
and sustain the RVI. In addition, we expect the RVI to initiate
limit cycles as mass builds up and the instability condition is
exceeded.   This is strongly   reminiscent of cataclysmic variables.

The  disk should
return to a stable, more quiescent state  when
sufficient radiation loss causes strong damping,  ultimately
stabilizing the  instability at some fraction of the critical column
density. This is suggestive of what
is observed in the water  maser nebula NGC 4258, which is
so accurately interpreted as a cool nebula in Keplerian orbit
around a SMBH (Claussen, Heiligman, \& Lo 1984; Henkel
\etal 1984).  Its column density is estimated to be
$\Sigma_{NGC} \simeq 10$ g cm$^{-2}$. We therefore interpret NGC
4258  as the residual disk left  over after forming the SMBH.
Such a disk should be self-gravitating, and indeed the drifts in
radio frequency lines are  interpreted (Bragg, Greenhill, \& Moran 2000) in
terms of spiral density waves.  This   supports  our
assumption that the strong shocks of the RVI should suppress the
SGI  until these shocks terminate.

\section{BH MASS AND VELOCITY DISPERSION DEPENDENCE}
\label{sec:bhmass}

In our simplified model of forming a flat rotation disk,  a
cloud with an initial radius $r_i$, total mass $M_0$, and spin
parameter
$\lambda_{eq}$  will collapse to
$r_0 = \lambda_{eq}^2 r_{i}$ with a column density
$\Sigma_0 = M_{0}/\pi r_0^2$, or
$\Sigma_0 = M_{0}/\pi (\lambda_{eq}^2 r_{i})^2$.
Then the total mass $M_{CCD}$ within the critical column
density $\Sigma_{CCD}$ of such a disk is
\begin{equation} M_{CCD}  =
\frac{M_{0}^2 }{\pi \Sigma_{CCD} (\lambda_{eq}^2 r_{i})^2}
\end{equation}

There should be a large dispersion in SMBH masses since the
variance to mean ratio of $\lambda$ from simulations is
$\sigma_{\lambda} \simeq 0.4$ (Warren \etal 1992),
and $M_{BH} \propto M_{0}^2 \lambda^{-4}$. We note that for a cloud
with uniform density and uniform rotation,  any sub-region
has the same $\lambda$ as the whole cloud, and so the collapse of
sub-regions will  follow equation (2).

The rotation velocity  of the disk
$v_{rot}$ is related to the disk column density by
$v_{rot}^2 = 2 \pi G \Sigma_0 r_0$,
    where G is the gravitational constant.  From equation (2), the mass
within a fixed column density
$\Sigma_{CCD}$ of such a disk is
\begin{equation}
M(>\Sigma_{CCD}) = {1\over 2 \pi
G^2} {1\over\Sigma_{CCD}} v_{rot}^4 \approx 3 \times 10^7
M_\odot~~,
\end{equation}
\noindent
if $v_{rot} = 200 \kms$ and $\Sigma_{CCD} = \Sigma_{RVI} = 100$ g
cm$^{-2}$.
   On very general grounds, we  expect $v_{rot}  \simeq \sqrt 3 \;
\sigma_c$, where $\sigma_c$ is the velocity dispersion of the bulge.
This  is because at the  radius where the spherical bulge meets the
flattened disk, $r_B$, the interior mass is identical for both
the bulge and the disk.  The extent to which the Keplerian velocity
at this point is greater than the flat rotation curve velocity   at
much larger radius is a measure of the increase in central mass
during mergers,
   $\Delta M_{merger} = M_{Flat} ((v_B / v_{Flat})^2 - 1) $.
In many cases this fractional increase in velocity near
the bulge is small, so the increase in mass is small.  Hence,
the mass inside $R_{CCD}$ becomes $M_{<CCD} \simeq 2.7 \cdot 10^8
M_{\odot}$, which agrees well  with observations by
Tremaine
\etal (2002), who discovered
$M_{BH,obs} = 1.35\times 10^8 (\sigma_c/200 \kms)^{4.02\pm
0.32}$.
   For galaxies that form their BH before mergers, we must use
$v_{rot}$  and assume that  the rotation curve is flat all
the way to the Rossby radius of $\sim$10 kpc when the galaxy is
very young.  Ferrarese (2002) has discovered a tight correlation
between
$\sigma_c$ and
$v_{rot}$:  $\log v_{rot}= (0.84 \pm 0.09) \log \sigma_c +
(0.55
\pm 0.19)$. Combined with Tremaine's observations, this gives
$M_{BH} = 1.82 \cdot 10^7 (v_{rot} / 200)^{4.79}$.
This is close to  our model, in which $ M_{BH} = 2.88
\times 10^7 ({v_{rot} / {200 {km}/s}})^4 M_\odot$.

\section{DISCUSSION AND CONCLUSIONS}
\label{sec:discuss}

We have proposed a simple model for the formation of a SMBH,
which is  synchronous with the formation of its host
galaxy from the gravitational collapse of a massive cloud. We rely
on two assumptions: (1) the initial local conservation of angular
momentum leading to a flat rotation disk and (2) the subsequent
initiation of the transport of angular momentum at a given column
density of the disk.   Together, these lead to a simple
explanation of the previously puzzling  mass of the SMBH
and  its mass-$\sigma$  relation. We have  related the
critical column density for the onset of angular momentum
transport with our work on a
large-scale instability in  accretion disks, the
RVI. Star formation
may follow exactly the same scenario in molecular clouds as SMBHs
in Ly$\alpha$ clouds. There is growing evidence  that
gravitational tidal torquing of "cores" in star formation regions
of molecular clouds  lead to the same values of
$\lambda_{av}
\simeq 0.05$  as in Ly$\alpha$  clouds  for galaxies.   Then
the mass of the cores within the  RVI limit becomes $\Sigma_{CCD}
\pi r_{CCD}^2 \sim  1 M_{\odot}$ for
$r_{core} \simeq 0.04$ pc  and $\lambda_{av} = 0.05$.

Gaseous disks formed at the
centers of galaxies are not exactly flat rotation  disks
even after gas  collisions following mergers.
Furthermore, mergers of binary SMBHs are expected to
evolve somewhat differently because of differing mass ratios and
impact parameters. These effects as well as the dispersion in
$\lambda_{av}$ will introduce  a finite dispersion in the
bulge velocity, as the $M_{BH}-\sigma_c$ relation indicates.
This cosmological explanation of the  $M_{BH}-\sigma$ relation
will be explored in a subsequent paper deriving the  SMBH mass
distribution  from the
$\lambda_{av}$  and  Press-Schechter distributions.

{\bf  Acknowledgements:}
The authors are deeply indebted to Ethan Vishniac as reviewer. One
of us, R. C.,  would like to thank Jeremy Goodman.
This research was performed under the
auspices of  the Department of Energy. It was supported by the
Laboratory Directed Research and Development Program  at Los Alamos
and in part by grants AST-02-06299. S. A. C and H. L. acknowledge the
hospitality of the Aspen Center for Physics while part of the
research was carried out during  the summer of 2002.


\begin{thebibliography}{DUM}

\bibitem[Adams, Graff, \& Richstone 2001]{agr01} Adams, F.C.,
Graff, D.S., \& Richstone, D.O. 2001, ApJ, 551, L31

\bibitem[Barnes \& Hernquist 1991]{bh91} Barnes, J., \&
Hernquist, L. 1991, ApJ, 370, L65

\bibitem[Begelman, Frank, \& Shlosman 1989]{bfs89}  Begelman,
MC., Frank, J., \& Shlosman, I. 1989,  "Theory of Accretion
Disks", eds. F. Meyer et al., Kluwer, p373.

\bibitem[Bragg \etal, 2000]{bra00} Bragg, A.E., Greenhill,
L.J., \& Moran J.M., 2000, \apj 535, 73

\bibitem[Bullock \etal 2001]{b01} Bullock, J.S., Dekel, A.,
Kolatt, T.S., Kravtsov, A.V., Klypin, A.A., Porciani, C., \&
Primack, J.R. 2001, ApJ, 555, 240

\bibitem[Claussen, Heiligman. \& Lo, 1984]{cla84}Claussen,
M.J.,  Heiligman, G.M., \& Lo, K.-Y,  1984 Nature, 310, 298

\bibitem[Colgate \& Buchler 2000]{cb00} Colgate, S.A., \&
Buchler, R.J. 2000, eds. Robert Buchler and Henry Kantrup,
Ann. N.Y. Acad. Sci. 898, 105.

\bibitem[Eisenstein \& Loeb 1995]{el95} Eisenstein, D.J., \&
Loeb, A. 1995, ApJ, 443, 11

\bibitem[Ferrarese 2002]{f02} Ferrarese, L. 2002,
\apj, 578, 90

\bibitem[Frank, King, \& Raine 1985]{fkr85}  Frank, J., King,
A.R., \& Raine, D.J. 1985, Accretion Power  in Astrophysics
(Cambridge: Cambridge University Press)

\bibitem[Gammie 2002]{g02} Gammie, C.F.,  2002, \apj 553, 174.

\bibitem[Goodman 2003]{g02} Goodman, J. 2003, MNRAS, 339, 937.

\bibitem[Henkel \etal 1984]{hen84} Henkel, C., Gu\"sten, R.,
Wilson, T.L., Bierman, P., Downes, D., \& Thum,C., 1984, A\&A,
141, L1

\bibitem[Li \etal 2000]{l00} Li, H., Finn, J.M.,
Lovelace, R.V.E. \& Colgate, S.A. 2000, ApJ, 533, 1023

\bibitem[Li \etal 2001]{l01} Li, H., Colgate, S.A., Wendroff,
B., \& Liska, R. 2001, ApJ, 551, 874

\bibitem[Loeb 1993]{l93} Loeb, A. 1993, ApJ, 403, 542

\bibitem[Lovelace \etal 1999]{l99} Lovelace, R.V.E, Li, H.,
Colgate, S.A. \& Nelson, A.F. 1999, ApJ, 513, 805

\bibitem[Mestel 1963]{m63} Mestel, L. 1963, MNRAS, 126, 553

\bibitem[Mihos \& Hernquist 1996]{mh96} Mihos, J.C., \&
Hernquist, L. 1996, ApJ, 464, 641

\bibitem[Ostriker 2000]{o00} Ostriker, J.P. 2000, Phys. Rev.
Lett., 84, 5258

\bibitem[Papaloizou \& Pringle 1984]{pp84}
Papaloizou, J.C.B., \& Pringle, J.E. 1984,
MNRAS, 208, 721

\bibitem[Peebles 1969]{p69} Peebles, P.J.E. 1969, ApJ, 155, 393

\bibitem[Pringle 1981]{pri81} Pringle, J.E.,  1981 Ann. Rev.
Astron. \& Astro Phys., 19, 137

\bibitem[Rees 1984]{r84} Rees, M.J. 1984, ARAA, 22, 471

\bibitem[Richstone \etal 1998]{r98} Richstone, D., \etal 1998,
Nature, 395, A14

\bibitem[Shakura \& Sunyaev(1973)]{shakura73} Shakura,~N.I.,
\& Sunyaev,~R.A. 1973, \aap, 24, 337

\bibitem[Shlosman, Begelman, \& Frank]{sbf90} Shlosman, I.,
Begelman, M.C., \& Frank, J. 1990, Nature, 345, 679

\bibitem[Steinmetz \& Bartelmann 1995]{sn95} Steinmetz, M., \&
Bartelmann, M. 1995, MNRAS, 272, 570

\bibitem[Tremaine \etal 2002]{t02} Tremaine, S.,
\etal 2002, ApJ, 574, 740

\bibitem[Umemura, Loeb, \& Turner 1993]{ult93} Umemura, M.,
Loeb, A., \& Turner, E.L. 1993, ApJ, 419, 459

\bibitem[Van den Bosch \etal 2002]{v02} Van den Bosch, F.C.,
Abel, T., Croft, R.A.C., Hernquist, L., \&  White, S.D.M. 2002,
ApJ, 576, 21

\bibitem[Warren \etal 1992]{w92} Warren, M.S., Quinn, P.J.,
Salmon, J.K., \& Zurek, W.H. 1992, ApJ, 399, 405

\end{thebibliography}
\end{document}